# UNVEILING THE MYSTERIES OF THE COSMOS: AN OVERVIEW OF RADIO ASTRONOMY AND ITS PROFOUND INSIGHTS

Manjuleshwar Panda[a], Yogesh Chandra[a*]

[a] Department of Physics, Government Post-Graduate College Bazpur, US Nagar, Uttarakhand 262401, INDIA
[*] Corresponding author: yepphysics@gmail.com

**ABSTRACT** — With its immensity and numerous mysteries waiting to be solved, the cosmos has always captivated humankind. A ground-breaking field that has given us a profound understanding of the mysteries of the cosmos is radio astronomy. This paper presents a comprehensive overview of radio astronomy, exploring its techniques, discoveries, and the profound insights it offers into celestial objects. Radio astronomy, which uses radio waves to analyse celestial phenomena, has completely changed how we think about the universe. This field has given us crucial information about the formation of stars, galaxies, and other celestial objects through the analysis of radio emissions. Radio astronomy has enabled researchers to study cosmic processes that are undetectable to the human eye by penetrating the furthest reaches of space. We explore radio astronomy techniques in this article, revealing how it can be used to see through interstellar dust and collect signals from the universe's furthest reaches. Pulsars, quasars, and cosmic microwave background radiation are significant discoveries that have helped astronomers understand dark matter and dark energy in great detail. We also look into how radio astronomy might be used in cosmology and astrophysics. In conclusion, radio astronomy has become a potent tool for solving the cosmos' riddles. Its capacity for the detection and analysis of radio emissions has produced a fundamental understanding of the beginnings and evolution of the universe. Radio astronomy continues to advance our understanding of the cosmos and arouses interest in additional cosmic research by shedding light on celestial objects that are invisible to the human eye.

## 1. INTRODUCTION

The cosmos has always piqued human attention because of its huge size and complicated web of celestial objects. Throughout history, astronomers have been devoted to unravelling the universe's mysteries and learning more about how it operates. Technological developments have created new areas for investigation, and radio astronomy is one such area that has fundamentally changed how we perceive the universe. Radio astronomy has shed light on the origins, evolution, and unobserved events of the cosmos by harnessing the power of radio waves.

The purpose of this research paper, "Unveiling the Mysteries of the Cosmos: An Overview of Radio Astronomy and Its Profound Insights," is to explore the field of radio astronomy by exploring its methods, findings, and the profound insights it has provided into the cosmos. By offering a thorough overview of its contributions to our understanding of celestial objects and cosmic processes, we will examine the enormous potential of radio astronomy as a strong instrument for uncovering the mysteries of the cosmos.

By enabling us to go beyond the limits of visible light, radio astronomy has completely changed how we view the cosmos. Scientists have uncovered hitherto unexplored facets of the cosmos by examining radio waves emitted by celestial objects. Radio waves have the distinct advantage of being able to pass through massive interstellar dust clouds that block out visible light, allowing for a clearer view of far-off astronomical objects. This quality has greatly increased our comprehension of the cosmos by allowing us to witness and study things that would otherwise remain concealed from our vision.

We will explore the methods used in radio astronomy in this paper, including the design and operation of radio telescopes, the nuances of radio wave detection, and the sophisticated data processing techniques used to extract useful information from the massive amounts of collected data. We'll go over some of the important discoveries made possible by radio astronomy, like the identification of pulsars, quasars, and cosmic microwave background radiation, each of which reveals a distinctive facet of the cosmos.

Beyond its practical contributions to human knowledge, radio astronomy has shed light on some of the most important issues surrounding the cosmos. The Big Bang theory has benefited greatly from the study of radio emissions, which has provided compelling evidence for the cosmic microwave background radiation, a byproduct of the early phases of the universe. We now know more about dark matter and dark energy, two enigmatic substances that govern the cosmos yet are difficult to measure directly thanks to radio astronomy. These deep realisations cast doubt on accepted beliefs and inspire additional investigation into the nature of the cosmos.

We will also look at the possible uses of radio astronomy in a variety of disciplines, such as astrophysics, cosmology, and astrobiology. Insights into stellar evolution, galactic dynamics, and the hunt for extraterrestrial intelligence have all advanced thanks to radio astronomy. Its influence extends far beyond our solar system, giving us a view into the outer reaches of the cosmos and inspiring awe and astonishment at the cosmic marvels that lay beyond our comprehension.

Finally, the purpose of this research study is to illuminate the amazing discipline of radio astronomy and its profound insights into the cosmos' mysteries. Radio astronomy has broadened our perspectives and transformed our understanding of the cosmos by utilising the power of radio waves. We wish to stimulate more investigation and pique the interest of future astronomers as they continue to solve the compelling mysteries that the cosmos presents by examining its methods, findings, and the profound insights it has offered.

2. **HISTORICAL OVERVIEW**

Humanity has always sought to understand the cosmos, but it wasn't until the development of radio astronomy that we were given a fresh understanding of it. The development of radio astronomy has been marked by many significant events, important discoveries, and the efforts of forward-thinking pioneers. We shall follow the development of radio astronomy in this historical overview, from its modest origins to its current prominence as a potent tool for solving cosmic puzzles.

The history of radio astronomy began in the 1930s when Karl Jansky, a Bell Laboratories engineer, unintentionally discovered radio waves coming from space. The discovery made by Jansky ushered in a new era of astronomical observations as well as the development of radio astronomy. Grote Reber, an amateur radio enthusiast who created the first purpose-made radio telescope in his backyard in the late 1930s, expanded on Jansky's work. Reber's innovative work produced the first radio image of the Milky Way and the mapping of radio sources. The science of radio astronomy was established after Jansky revealed his discovery in April 1933 at a meeting in Washington, D.C. **[1]**. The revolutionary work of Jansky and Reber established the groundwork for the rapid development of radio astronomy. Their discoveries provided astronomers with whole new perspectives on how to observe and comprehend the cosmos, allowing them to investigate celestial objects and processes that had previously eluded optical telescopes. Radio astronomy has developed into a potent and critical instrument for astronomers throughout the years, offering crucial insights into the cosmos and assisting in significant scientific advancements.

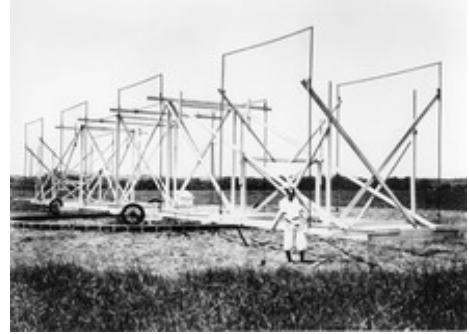

**Fig 1.1: The first radio telescope in the world, built by Karl Jansky in Holmdel, New Jersey, in the early 1930s, was used to find radio signals coming from the Milky Way.**

Radio astronomy, however, really took off during and following World War II. Many engineers and scientists who had worked on radar systems during the war began to focus on radio astronomy. By using repurposed military radar equipment to monitor cosmic radio emissions, Bernard Lovell and his team at Australia's Jodrell Bank Observatory made important advancements in radio astronomy. Their work was successful when radio signals from the Sun and other celestial bodies were discovered. The advancement of increasingly advanced radio telescopes in the 1950s accelerated the field. Notably, a noteworthy accomplishment was the building of the 76-meter Mark I radio telescope at Jodrell Bank Observatory **[2].** This enormous device allowed for accurate positioning measurements of radio emitters and grew to become the largest fully steerable radio telescope in the entire globe. This period saw significant advances in our knowledge of the energetic phenomena of the cosmos, including the identification of quasars and the discovery of the first radio galaxy, Cygnus A. With the introduction of aperture synthesis methods, radio astronomy made a significant advancement in the 1960s. The Cavendish Laboratory's Martin Ryle and Antony Hewish invented the idea of interferometry, which combines a number of smaller telescopes to create a bigger virtual telescope. This discovery paved the way for the development of the Cambridge Interferometer, a ground-breaking device with previously unheard-of sensitivity and resolution. Ryle and his team conducted comprehensive surveys of the radio sky using this new capability, finding many radio sources and analysing the structure of galaxies.

The development of bigger, more accurate radio telescopes marked the subsequent decades' progress in radio astronomy. Completed in 1963, the Arecibo Observatory in Puerto Rico had a 305-meter spherical dish that offered unmatched sensitivity for researching pulsars, radio galaxies, and the hunt for extraterrestrial intelligence. The installation of radio interferometer arrays, such the Very Large Array (VLA) in the United States and the Australia Telescope Compact Array (ATCA), also made it possible to image cosmic occurrences in high detail. The creation of next-generation devices and partnerships has elevated radio astronomy in recent years. In-depth studies of star formation, protoplanetary discs, and the early cosmos are now possible because to the Atacama Large Millimeter/submillimeter Array (ALMA), a project of worldwide cooperation. Additionally, with its unmatched sensitivity and survey capabilities, the Square Kilometre Array (SKA), a bold worldwide endeavour, promises to revolutionise radio astronomy.

Our understanding of the cosmos has surely changed as a result of the advancement of radio telescopes and observational methods. Radio astronomy has uncovered a multitude of cosmic mysteries, from the ground-breaking work of Jansky and Reber through the ground-breaking accomplishments of Lovell, Ryle, and their contemporaries. Radio astronomers have advanced our understanding of the universe by embracing cutting-edge technology and novel methods. As a result, they have inspired future generations to continue exploring the deep mysteries of the cosmos.

3. FUNDAMENTALS OF RADIO ASTRONOMY

Understanding the foundations of this fascinating science is crucial to fully appreciating the tremendous insights radio astronomy has offered. The fundamental principles that underpin radio astronomy are examined in this section. Therefore, we'll look into electromagnetic radiation, radio waves, the electromagnetic spectrum, and how radio waves interact with celestial objects.

The fundamental building block of our understanding of the cosmos is electromagnetic radiation. It includes a wide variety of energy that moves as waves, each of which has a specific wavelength and frequency. Radio waves are a subset of electromagnetic energy that are at the core of radio astronomy. These waves have low frequencies and rather long wavelengths, varying from a few millimetres to several metres. Electromagnetic radiation is a form of energy that propagates as both electrical and magnetic waves traveling in packets of energy called photons. It contains the waves of the electromagnetic field, which propagate through space and carry momentum and electromagnetic radiant energy [3]. Electromagnetic waves are synchronised oscillations of the electric and magnetic fields that make up electromagnetic radiation in the classical sense. Different electromagnetic spectrum wavelengths are produced depending on the frequency of oscillation. Since photons are uncharged elementary particles with zero rest mass and are the electromagnetic field's quanta, which are responsible for all electromagnetic interactions, this alternative interpretation of EMR is based on quantum physics. When a charged particle, like an electron, changes its velocity—that is, when it is accelerated or decelerated—electromagnetic radiation is created. The charged particle is responsible for losing the energy of the electromagnetic radiation that is thus created.

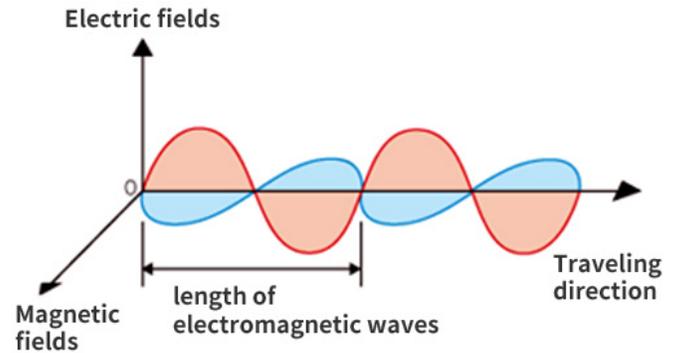

Fig 1.2: Electromagnetic Wave

A wide range of electromagnetic energy, from extremely long radio waves to extremely brief gamma rays, travels in waves. The name of this spectrum is electromagnetic spectrum. All electromagnetic waves are included in the electromagnetic spectrum. Radio waves, microwaves, infrared, visible light, ultraviolet, X-rays, and gamma rays are all included in the electromagnetic spectrum, which gives a thorough overview of the whole range of electromagnetic radiation. A discrete region with its own characteristics and interactions with matter is represented by each segment of the spectrum. Because they may pass through barriers like interstellar dust clouds that can obscure visible light studies, radio waves, which are at the lower end of the electromagnetic spectrum, are particularly important for radio astronomy.

- **Radio:** Your radio picks up radio waves broadcast by radio stations and transmits your favourite music to you. Stars and space gases can also emit radio waves.

- **Microwave:** Microwave radiation not only quickly cooks popcorn, but it is also utilised by astronomers to study nearby galaxies' structures.

- **Infrared:** Night vision goggles can detect the infrared light that is released by hot objects including human skin. Infrared light allows us to map the dust between stars in space.

- **Visible:** Visible light is seen by our eyes. Visible light is produced by stars, fireflies, and lightbulbs.

- **Ultraviolet:** Skin tans and burns are caused by ultraviolet light, which the Sun emits. In space, "hot" objects also release UV light.

- **X-rays:** X-rays are used by dentists to image your teeth and by airport security to inspect your bag. Additionally, the universe's hot gases release X-rays.

- **Gamma-ray:** Gamma-ray imaging allows medical professionals to view into your body. The Universe is the largest gamma-ray generator known to man.

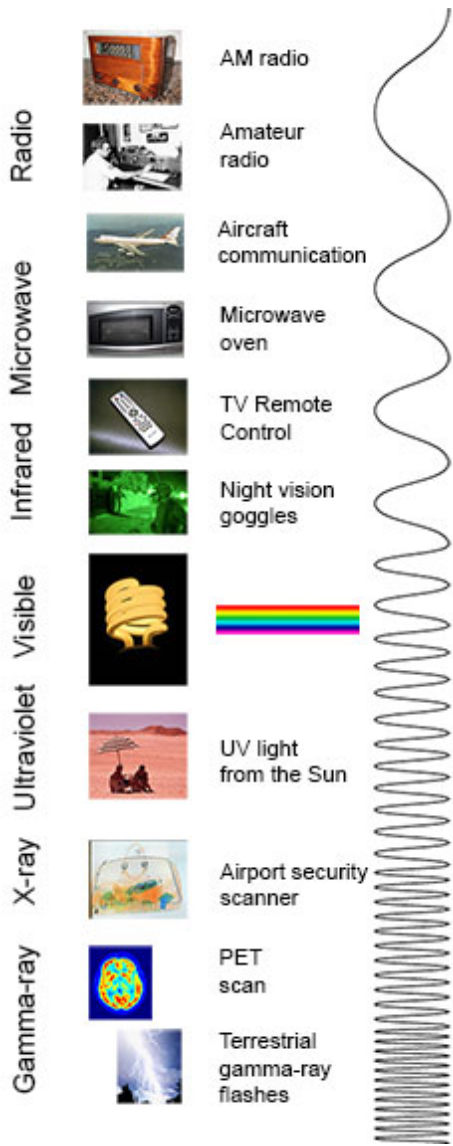

**Fig 1.3: The electromagnetic spectrum from lowest energy/longest wavelength (at the top) to highest energy/shortest wavelength (at the bottom). (Credit: NASA's Imagine the Universe)**

Specific intervals within the electromagnetic spectrum's radio wave region are known as radio frequency bands. These bands, which may be categorised and studied using their frequency range, offer a foundation for doing so. The metre bands (from decameter to millimetre wavelengths) and the microwave bands (from centimetre to millimetre wavelengths) are the radio frequency ranges that are most frequently employed in radio astronomy. Each band permits astronomers to study the cosmos at various levels of detail and provides distinct insights into particular astrophysical processes.

Radio astronomy observations are based on the interaction of radio waves with celestial objects. Through a variety of processes, including thermal radiation, synchrotron radiation, and maser emissions, celestial objects release radio waves. These things interact differently with radio waves, which reveals important details about their makeup, temperature, velocity, and magnetic fields. Thermal radiation is a prime example of this interaction. Because they are naturally hot, celestial objects like galaxies and stars generate radio waves. Astronomers can learn about the temperature and physical characteristics of these objects by observing the strength and spectrum of these radio waves. Furthermore, high-energy charged particles, like electrons, spiral along magnetic field lines, which is how synchrotron radiation is produced. The study of cosmic occurrences like supernova remnants and active galactic nuclei is made possible by the phenomenon's distinctive radio emissions. Additionally, maser emissions provide a distinctive window into areas of high activity. The acronym **"Maser"** [4] stands for "microwave amplification by stimulated emission of radiation." Masers amplify and emit microwave and radio waves in specific astrophysical settings, such as molecular clouds, in a highly focused and coherent manner. Astronomers can examine the activities taking place in these locations, such as star formation and the presence of complex chemicals, by spotting and examining maser emissions.

Overall electromagnetic radiation, the electromagnetic spectrum, and the interaction of radio waves with celestial objects are the core concepts of radio astronomy. Astronomers have learned a great deal about the universe by using radio waves, which are part of the electromagnetic spectrum. Understanding these fundamental ideas paves the way for understanding the amazing discoveries and insights that radio astronomy has produced, allowing us to solve cosmic mysteries and increase our understanding of the cosmos.

4. **RADIO TELESCOPE TECHNOLOGY**

The radio telescope technology acts as our steadfast guide as we travel further into the centre of the cosmos, staring into the immense expanse of the universe where light is illusive but radio waves reveal the most profound mysteries of celestial objects. This Part will dig into the fascinating world of radio telescopes, examining the various types, designs, and functions of these instruments as well as the amazing developments that have taken radio astronomy to new heights.

**4.1 Radio Telescopes: Unraveling the Invisible**

The purpose of radio telescopes, which are astronomical devices, is to identify and investigate radio waves emitted by celestial objects and events. These radio waves carry crucial information about distant objects and events, allowing us to uncover the mysteries of the cosmos in ways that traditional optical telescopes cannot achieve **[5].** While radio telescopes record radio frequencies, which have longer wavelengths, optical telescopes focus on visible light. A radio telescope is a specialised antenna and radio receiver used to find radio waves coming from astronomical radio sources in the sky. The primary observational tool used in radio astronomy, which investigates the radio frequency region of the electromagnetic spectrum radiated by celestial objects, is the radio telescope.

**4.2 Different Types of Radio Telescopes**

**4.2.1 Single-Dish Radio Telescopes**

The most basic and conventional type of radio telescope is the single-dish model. It is made up of a sizable reflector in the form of a dish that gathers incoming radio waves and concentrates them onto a receiver at the focal point. The telescope's sensitivity and resolving power depend on the dish's size. Renowned examples include the Arecibo Observatory in Puerto Rico (sadly decommissioned in 2020) and the Parkes Observatory in Australia, which played a crucial role in receiving the first signals from the Apollo 11 moon landing **[6].**

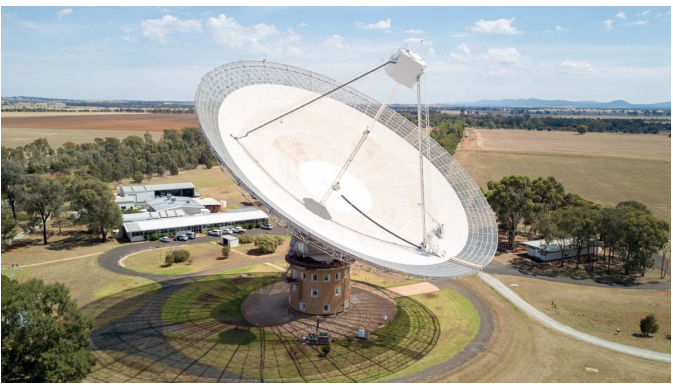

**Fig 1.4: The Parkes 64m Radio Telescope**

**4.2.2 Radio Interferometers**

The development of radio interferometers by radio astronomers allowed them to overcome the constraints of single-dish telescopes. These devices are intended to operate as an array and are made up of numerous smaller radio antennas dispersed over large distances. Interferometers produce substantially higher resolution and sensitivity by combining the signals from each antenna, similar to building a virtual telescope with a size equal to the maximum distance between the antennas. The Very Large Array (VLA) in the United States and the Atacama Large Millimeter/submillimeter Array (ALMA) in Chile are two prominent interferometric arrays.

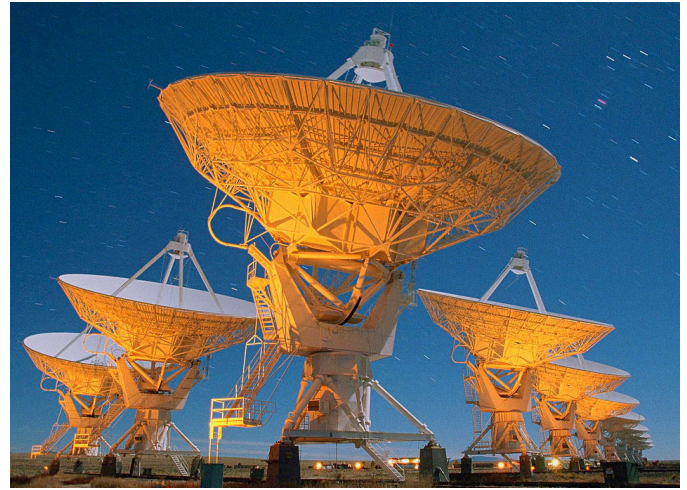

**Fig 1.5: Very Large Array (VLA) in Socorro, New Mexico**

**4.3 Design and Functioning of a Typical Radio Telescope**

The reflector or antenna, the receiver, and the data processing system make up the core of a radio telescope's fundamental construction.

- **The Reflector:** Incoming radio waves are captured and focused by the reflector, which is its main job. Reflectors come in a wide range of sizes and shapes, with parabolic dishes being the most popular. The sensitivity and capacity to find weak signals increase with the size of the dish.
- **The Receiver:** The concentrated radio waves are transformed into electrical signals that may be analysed and processed by the receiver. In order to reduce noise and improve the sensitivity of the telescope, modern receivers use very sensitive detectors such superconducting materials or low-noise amplifiers.
- **The Data Processing System:** To obtain useful astronomical data from the receiver's data collection, processing and analysis are required. As a result, precise radio maps and spectra of the detected celestial objects are produced using sophisticated computer systems and algorithms that clean and interpret the signals.

A big dish-shaped reflector, which concentrates the incoming waves onto a receiver at the focal point, is the key component of a radio telescope's mechanism for collecting radio waves from astronomical objects. The radio waves are transformed into electrical signals by the receiver, and these electrical signals are then processed by a sophisticated data processing system, requiring intricate algorithms and computer systems, to provide precise radio maps and spectra of the seen celestial objects.

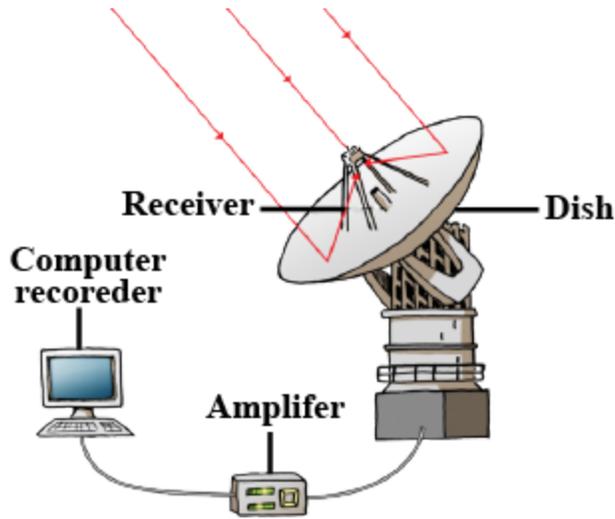

**Fig 1.6: Radio Telescope System**

The parabolic surface of some radio telescopes is positioned equatorially, aligned with the axis of rotation of the Earth. By moving the antenna along a single axis, this configuration enables the telescope to track a place in the sky while Earth rotates. Equatorial mounts, however, are difficult and expensive to build. Modern radio telescopes have digital computers that effectively control the azimuth and elevation axes of the telescope, allowing for accurate monitoring of radio emitters across the sky.

**4.4 Modern Advancements in Radio Telescope Technology**

Recent years have witnessed remarkable advancements in radio telescope technology, fueling a new era of discovery in radio astronomy **[7].**

- **Aperture Synthesis:** Radio astronomers can synthesise a large virtual aperture using this method, which is utilised in interferometers, by mixing signals from various antennas. It has substantially increased the radio telescopes' resolving power, allowing us to observe detailed details in far-off galaxies and star-forming areas.
- **Software-Defined Radio (SDR):** Instrumentation for radio telescopes has undergone a revolution because to SDR technology, which digitises signals right at the antennas. This adaptability enables more complex data processing, simpler calibration, and effective use of the telescope in a variety of observation modes.
- **Big Data and Machine Learning:** As radio telescopes generate vast amounts of data, the application of big data and machine learning techniques has become indispensable. These methods aid in the analysis, classification, and extraction of relevant information from the deluge of data received by radio telescopes.

Modern astronomy is led by radio telescope technology, which reveals the mysteries of the cosmos concealed in radio waves. Their designs and capabilities have greatly changed, from simple single-dish telescopes to complex interferometric arrays, enabling us to explore the universe in unheard-of ways. Radio astronomy will continue to provide significant insights into the nature and origins of our universe with continuous improvements and upcoming developments.

## 5. RADIO ASTRONOMY OBSERVATIONS

By seeing into the depths of space through the radio waves emitted by celestial objects, radio astronomy, an amazing field of study, has allowed humanity to discover the mysteries of the cosmos. In this chapter, we explore the fascinating methods used in radio astronomy investigations, the complexities of data collection and processing, and the difficulties encountered in this search for cosmic comprehension.

### 5.1 Techniques Used in Radio Astronomy Observations

#### 5.1.1 Continuum Imaging

The ability to create high-resolution photographs of the radio sky via continuum imaging is one of the core methods used in radio astronomy. By capturing the fluctuations in radio source intensity, this method can shed light on the composition and emission processes of astronomical objects. Fourier transform methods are used in the mathematical framework of continuum imaging to recreate images from interferometric data **[8].** The fundamental measurement derived from interferometric observations is the visibility function ($V(u, v)$), which represents the correlation between the signals received at two telescopes separated by a vector $(u, v)$ in the plane of the sky. The radio sky's brightness distribution is obtained from the visibility function's two-dimensional Fourier transform, designated as $I(l, m)$. The following is an expression for the relationship between the visibility function and the brightness distribution:

$$I(l,m) = \iint V(u,v) \cdot e^{2\pi i(ul+vm)} \, du \, dv$$

Where:

- **I(l, m)** is the brightness distribution at coordinates (l, m) in the sky,
- **V(u, v)** is the visibility function at spatial frequency coordinates (u, v),
- **(u, v)** are the spatial frequency coordinates in the Fourier plane, and
- **e^{2\pi i (ul + vm)}** is the complex exponential term that accounts for the phase shift due to the position (l, m) in the sky.

We may see the emission sources and learn about their structures by applying the inverse Fourier transform to the visibility data to rebuild the radio sky image. This mathematical illustration highlights the fundamentals of continuous imaging in radio astronomy, where the transformation of the visibility function and brightness distribution is essential for solving the puzzles of celestial objects **[9].**

**5.1.2 Spectroscopy**

Finding and analysing the spectral lines generated by diverse astronomical sources requires the use of radio spectroscopy. These spectral lines' Doppler shift reveals important details on the speed and characteristics of far-off galaxies, interstellar gas, and even elusive dark matter **[10].** Our understanding of the universe is enriched by the analysis of these spectra, which reveals the structure and movement of celestial objects.

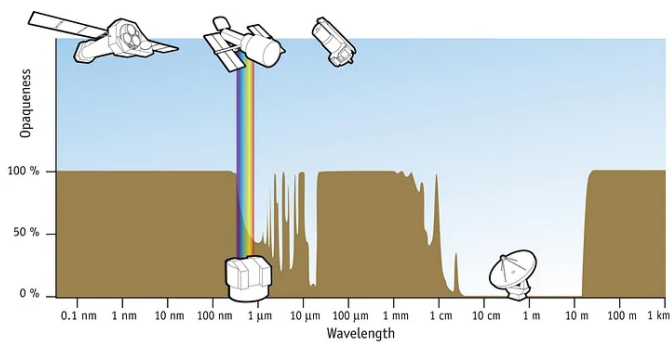

**Fig 1.7: In the EM spectrum, radio waves, X-rays, and visible light have the lowest opacity or opaqueness. (Source:Wikimedia)**

In radio spectroscopy, two different types of radio-spectral lines are frequently employed:

- The H I 21 cm Line is used to determine the atomic gas mass of high redshift galaxies and is conceivably the most widely used line in radio astronomy.
- The CII-158 μm line for ionized carbon is used to study high redshift disc galaxies.

Here are some of the explanations for why radio spectroscopy is so crucial and why you should adore it:

> As was already mentioned, radio waves and visible light can both be detected terrestrially since they have the least air opacity.

> Both optical and radio observations can be made from Earth with large telescopes, but radio is more useful since the atmosphere is less opaque. The radio sky: frequency of nearly 10 MHz — 1THz; wavelength of around 30m to 0.3mm. On the other hand, the optical sky: wavelength around 3000 Å — 10,000 Å.

> Anyone can do radio spectrography from any observatory, which is possible thanks to the world's observatories.

**5.1.3 Polarisation**

Polarisation measurements in radio astronomy allow researchers to understand the magnetic fields and physical conditions prevailing within astronomical sources. The Stokes parameters are used to quantify polarisation, which reveals intricate details about the source's magnetic fields and the processes that lead to polarised radiation **[11].**

**5.2 Data Acquisition and Processing Methods**

**5.2.1 Radio Telescopes and Interferometry**

Aperture synthesis is one example of the cutting-edge technology used by contemporary radio telescopes to improve sensitivity and produce high-resolution imagery. In order to create a virtual telescope with a diameter equal to the greatest possible spacing between individual antennas, interferometric arrays integrate the signals from various telescopes. This method dramatically boosts resolution, giving a crisper picture of far-off astronomical objects **[12].**

**5.2.2 Signal Processing and Calibration**

Telescopes pick up weak radio signals that are frequently distorted by numerous types of noise. In order to extract useful information from the raw data, sophisticated signal processing techniques like the Fast Fourier Transform (FFT) and CLEAN algorithm are used. Calibration procedures correct for instrumental effects, atmospheric distortions, and other imperfections, ensuring the accuracy and reliability of the final results **[13].**

**5.3 Challenges Faced in Radio Astronomy Observations and Their Solutions**

### 5.3.1 Radio Frequency Interference (RFI)

Radio astronomical observations may become contaminated by RFI from terrestrial sources like satellites and communication equipment. Radio observatories are strategically placed in radio-quiet areas, and when processing data **[14]**, sophisticated RFI mitigation methods are used.

### 5.3.2 Atmospheric Absorption and Ionospheric Distortions

Radio waves can be absorbed and distorted by the atmosphere of the Earth, especially at some frequencies. Astronomers frequently watch at higher altitudes and in frequency ranges with less atmospheric influence to combat this. In order to account any residual atmospheric distortions **[15]**, advanced correction methods are also used.

### 5.3.3 Data Volume and Processing Speed

Large volumes of data are generated by radio astronomy, especially with interferometric arrays. Such data are extremely computationally intensive to store and process. In order to deal with the data flood **[16]**, radio observatories deploy high-performance computing clusters and adopt data reduction techniques.

## 6. COSMIC MICROWAVE BACKGROUND (CMB) RADIATION

Deep understanding of the cosmos' creation, development, and structure can be gained from studying the cosmic microwave background (CMB) radiation, a soft glow that permeates the whole universe. This chapter explores the definition of the CMB, its significance, the ground-breaking findings and insights that have been obtained from CMB observations **[17]**, and the crucial role that radio telescopes have played in revealing the secrets of the CMB.

### 6.1 Explanation of the CMB and Its Significance

The remnants of the Big Bang, which took place roughly 13.8 billion years ago, are now known as the **cosmic microwave background radiation.** The initial plasma of photons and matter started to dissociate as the cosmos expanded and cooled, allowing the photons to move freely through space. The universe's expansion has cooled and stretched these old photons, causing them to become the microwave wavelengths they are today. The electromagnetic spectrum's microwave region is where the background radiation is found.

A snapshot of the early universe when it first became transparent to light is provided by the CMB, which is of utmost significance to cosmology. Astronomers can learn vital details about the age, make-up, shape, and pace of expansion of the universe by investigating the characteristics of the CMB. It is a potent tool for comprehending the basic laws of the universe and supporting the Big Bang hypothesis.

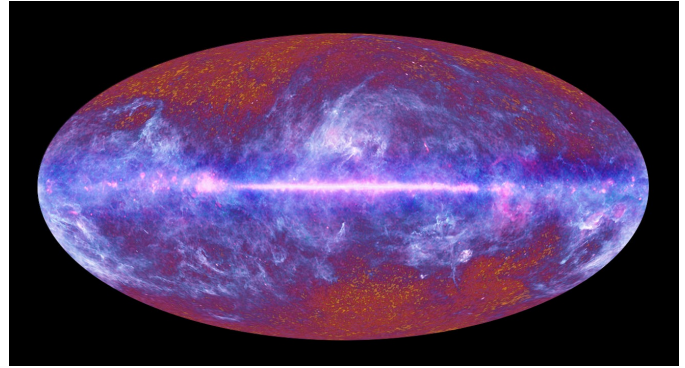

**Fig 1.8: The Planck telescope simultaneously recorded two photos that span nearly the entire 13.7 billion year history of the cosmos in this single all-sky image. (Credit: ESA)**

Pioneering space missions have concentrated on researching the cosmic microwave background (CMB) radiation in their attempt to unravel the universe' mysteries. The 1989-launched NASA COBE instrument showed the black-body spectrum of the CMB at 2.73 Kelvin as well as minuscule temperature variations in the sky. When WMAP was developed in 2001, it probed these fluctuations more deeply and discovered the traces of early density fluctuations that later affected the large-scale structure of the cosmos. An even more thorough perspective was supplied by ESA's Planck, which was launched in 2009 and precisely analysed the CMB and foreground sources to improve our understanding of cosmology. These missions have established the cosmological standard model and looked into the intriguing prospect of new physics beyond it.

### 6.2 Discoveries and Insights from CMB Observations

#### 6.2.1 Anisotropies and Primordial Density Fluctuations

A remarkable finding from the investigation of the cosmic microwave background (CMB) is the distribution of tiny temperature variations, or anisotropies, over the celestial sphere. These minuscule temperature changes are the remains of the early density fluctuations that appeared at the creation of the cosmos. They were crucial in determining the structure of the cosmos and eventually gave rise to the galaxies, clusters, and superclusters that we see today. Astronomers learn crucial information about the conditions in the early universe by examining the patterns of these anisotropies, which aids in solving its riddles and validating cosmological models. A wealth of knowledge is revealed by the careful study of the CMB data in conjunction with theoretical frameworks, which reveals the mysteries of the universe's evolution and the distribution of matter on cosmic scales. These findings have solidified the CMB's status as a priceless cosmic map that leads us through the complex evolution of our cosmos.

### 6.2.2 Cosmic Inflation

The strong evidence for cosmic inflation is one of the most important findings from CMB measurements. The cosmos experienced inflation, a quick exponential expansion, right after the Big Bang. The large-scale structure of the universe and the observable uniformity of the CMB are explained by this process. The inflationary paradigm is supported by the CMB data, which sheds light on the early universe when examined with theoretical models.

### 6.3 The Role of Radio Telescopes in CMB Research

Radio telescopes are essential for CMB studies because they can measure the temperature variations in the CMB radiation with great accuracy. These telescopes have extremely sensitive sensors that can pick up the weak microwave emissions from the CMB. Astronomers can map the CMB anisotropies with unmatched precision using interferometric radio telescope arrays, enabling in-depth investigations of the characteristics of the early cosmos. The CMB data [18] can be used to extract useful cosmological information thanks to radio telescopes and cutting-edge data analysis technologies like maximum likelihood approaches and Gibbs sampling.

A key finding in radio astronomy, the cosmic microwave background radiation provides priceless insights into the origins and development of the universe. The CMB's observations and research continue to influence our understanding of the universe, and radio telescopes are essential to this cosmic quest.

## 7. GALACTIC RADIO ASTRONOMY

The study of galactic radio astronomy provides a window into the secrets of the Milky Way galaxy, which is our own galaxy. We set out on a voyage through the fascinating radio wave study of our galaxy in this chapter. Radio astronomy offers unique insights into the cosmic activities taking place in our galaxy, from mapping neutral hydrogen (HI) to finding and analysing pulsars, supernova remnants, and other galactic occurrences.

### 7.1 Study of Our Own Milky Way Galaxy Using Radio Waves

Astronomers trying to learn more about the Milky Way galaxy face particular difficulties due to its huge size. However, radio waves have a distinct advantage since they can cut through thick interstellar dust clouds and expose celestial objects that are normally hidden from view by optical wavelengths. The wide variety of objects in our galaxy are revealed by galactic radio astronomy, from gigantic black holes hiding at the galactic nucleus to stellar nurseries and star-forming regions.

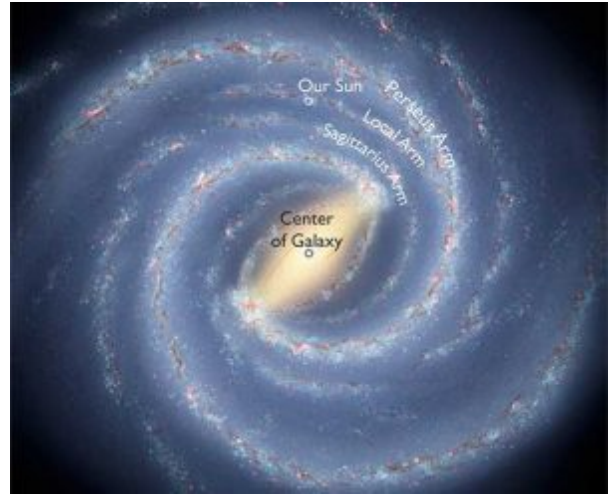

**Fig 1.9: The massive gas clouds surrounding the Sun appear to move in a manner that is comparable to that of the clouds in the neighbouring gigantic arm of Perseus, according to radio data from the VLBA. Our neighbourhoods span the galaxy. (Credit: NRAO)**

Our solar system is embedded within the galaxy, and much of the light released by the galaxy's stars is blocked off by interstellar dust and gas, making it extremely challenging to study our own galaxy with light waves. This issue is better resolved by radio astronomy since radio waves may pass through dust and gas in the way and produce pictures of the galaxies' internal architecture. The **8-inch (21-cm) line** in the radio spectrum, which is emitted by hydrogen atoms, is particularly significant in these research. The distribution of interstellar gas and dust inside the galaxy can be determined using the 8-inch (21-cm) line. Radio astronomers have another useful tool for examining the structure of our galaxy in the form of radio emission from molecules in the interstellar plasma.

### 7.2 Mapping of Neutral Hydrogen (HI) in the Galaxy

The Milky Way's neutral hydrogen (HI) gas distribution has been extensively mapped using radio telescopes. The interstellar medium's primary constituent, HI, offers crucial hints about the galaxy's structure, rotation, and dynamics. Our galaxy's spiral arms, galactic bars, and other structural features may all be traced using the 21-centimeter emission line of HI, providing a thorough insight of the complex structure of our galaxy.

## 7.3 Detection and Analysis of Pulsars, Supernova Remnants, and Other Galactic Phenomena

The discovery and investigation of numerous spectacular events inside the Milky Way is made possible by radio astronomy. Pulsars are neutron stars that rotate quickly and release beams of radio waves. They provide information on extreme physics and act as celestial beacons for the entire galaxy [19]. Massive star explosion leftovers known as supernova remnants offer important information on stellar evolution and the enrichment of the interstellar medium with heavy metals. Radio measurements have also added to our understanding of the galactic environment by shedding light on cosmic rays, masers, and galactic magnetic fields.

The Milky Way's celestial glories are still being revealed by galactic radio astronomy, which fosters a close relationship with our cosmic home and adds to the richer tapestry of radio astronomy's profound discoveries.

## 8. EXTRAGALACTIC RADIO ASTRONOMY

By extending our investigation outside the Milky Way, extragalactic radio astronomy enables us to solve the cosmic puzzles of far-off galaxies and quasars. We dig into the fascinating discoveries made possible by observing these heavenly objects using radio waves in this chapter. Extragalactic radio astronomy offers remarkable insights into the universe outside of our own galaxy, from radio jets and active galactic nuclei (AGN) to cosmic evolution and large-scale structural investigations.

## 8.1 Investigation of Distant Galaxies and Quasars Using Radio Waves

In order to better understand the physical characteristics and evolution of far-off galaxies and quasars, radio telescopes are effective study tools. Astronomers can study these cosmic objects' star formation rates, stellar composition, and potentially locate elusive supermassive black holes at their centres by observing their radio emissions. Our knowledge of the history and diversity of the universe is enriched by extragalactic radio astronomy, which reveals the cosmic drama taking place in distant galaxies.

## 8.2 Radio Jets, Active Galactic Nuclei (AGN), and Radio Galaxies

Few phenomena in the field of extragalactic radio astronomy are as intriguing as the mysterious radio jets and the massive Active Galactic Nuclei (AGN) centres. These cosmic engines, propelled by supermassive black holes, unleash magnificent bursts of energy, giving rise to some of the universe's [20] brightest objects. A supermassive black hole, millions to billions of times as large as our Sun, is located at the centre of AGN. The **Active galactic nucleus (AGN)** is a tiny area at the galaxy's core that emits a tremendous amount of energy as radio, optical, X-ray, or gamma radiation or high-speed particle jets. A dazzling accretion disc develops around the black hole as the surrounding material is irrevocably sucked into its gravitational pull. Huge amounts of energy over the electromagnetic spectrum, including powerful radio broadcasts, are released as a result of the enormous gravitational forces compressing and heating the gas inside the disc to extreme temperatures. These radio waves are the result of charged particles spiralling at high speeds in a strong magnetic field close to the event horizon of a black hole.

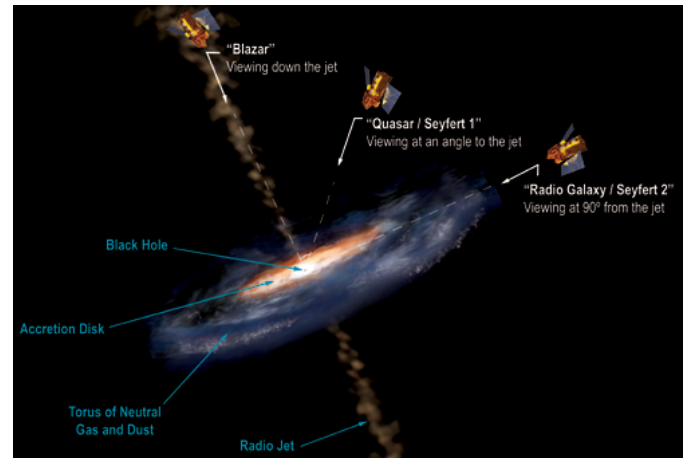

**Fig 1.10: Active Galactic Nuclei (Credit: Aurore Simonnet, Sonoma State University)**

AGNs have three crucial characteristics: a black hole, an accretion disc, and jets. The driving force behind the entire process is the black holes at the center of an AGN. AGN's breathtaking radio jets are one of their distinctive characteristics. These jets are two extremely fast plasma outflows that emerge from the area around the supermassive black hole at speeds that are almost as fast as light. Although the exact cause of these jets is still unknown, it is most likely due to the strong magnetic fields and relativistic processes close to the black hole. These jets create massive channels through space as they move out into the intergalactic medium, producing radio waves that can travel millions of light-years beyond the host galaxy.

An exclusive subclass of AGN known as radio galaxies produce intense, long-range radio emissions. These galaxies frequently inhabit crowded areas like galaxy clusters, where interactions and mergers between galaxies are frequent occurrences. Intense radio jet activity is caused by the accretion of matter onto supermassive black holes, which is triggered by galaxy collisions. As a result, radio galaxies transform into cosmic energy beacons that illuminate their environs and have an impact on the dynamics of their local cosmic communities. The distinctions between the various kinds of active galaxies and typical galaxies are shown in the following table:

| Galaxy Type | AGN | Strong Radio | Jets |
|---|---|---|---|
| Normal | No | No | No |
| Starburst | No | Some | No |
| Seyfert I | Yes | Few | No |
| Seyfert II | Yes | Few | Yes |
| Quasar | Yes | Some | Some |
| Blazar | Yes | Yes | Yes |
| BL Lac | Yes | Yes | Yes |
| OVV | Yes | Yes | Yes |
| Radio Galaxy | Yes | Yes | Yes |

**Table I: Features - Normal vs Active Galaxies**

The interconnectedness of supermassive black holes and their host galaxies can be better understood through researching radio jets, AGN, and radio galaxies. The evolution of the galaxy, star formation, and the dispersion of matter on cosmic scales are all significantly influenced by these phenomena. Additionally, radio jets contribute significantly to feedback processes by releasing energy into their surroundings, which influences the expansion of nearby galaxies and the universe's evolution.

**8.3 Cosmic Evolution and Large-Scale Structure Studies**

Understanding the history of galaxies and the large-scale structure of the cosmos [21] requires the use of extragalactic radio astronomy 2. Astronomers can determine the distribution of dark matter, the growth of cosmic structures over billions of years, and the expansion of the universe by looking at the cosmic web of galaxy clusters and cosmic gaps. These investigations provide information on the underlying mechanisms that have shaped the universe during the universe's long cosmic history.

The study of extragalactic radio astronomy broadens our understanding of the cosmos by giving us fascinating insights into the immense fabric of galaxies, quasars, and cosmic structures, which piques our interest and motivates new discoveries.

### 9. TRANSIENT RADIO ASTRONOMY

Transient radio astronomy emerges as an enthralling conductor in the cosmic symphony of the ever-evolving universe, directing the observation and analysis of transient astronomical occurrences with cutting-edge radio telescopes. This chapter sets off on an exciting voyage into the world of fleeting events, revealing their mysterious nature and significant implications for our comprehension of the cosmos. Transient radio astronomy reveals a tapestry of cosmic marvels that shed light on the secrets of the cosmos, from cosmic explosions like fast radio bursts (FRBs) and gamma-ray bursts (GRBs) to the precise timing of pulsars.

**9.1 Detecting and Studying Transient Events using Radio Telescopes**

Our understanding of the cosmos is revolutionised by transient radio astronomy, which captures the splendour of fleeting events that arise and disappear in cosmic winks. Radio telescopes [22] operate as attentive cosmic sleuths, recording these transient occurrences thanks to their exceptional sensitivity and quick response times. In addition to a bizarre object detected in 2018, astronomers have discovered two other objects that together form a new class of cosmic explosions. The new form of explosion has different features from each but does have some traits with supernova explosions of large stars and with explosions that produce gamma-ray bursts (GRBs).

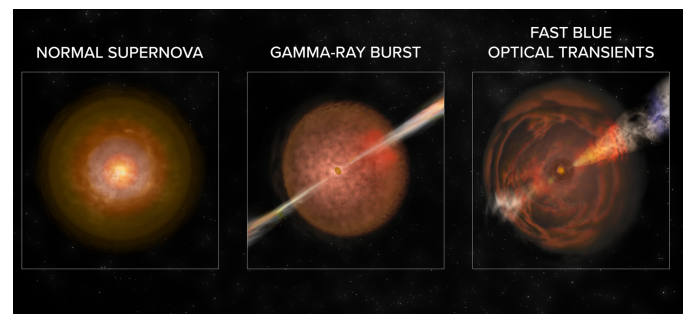

**Fig 1.11: The contrast in phenomena between a 'ordinary' core-collapse supernova explosion, an explosion producing a gamma-ray burst, and an explosion producing a Fast Blue Optical Transient are shown in an artist's conception. (Credit: Bill Saxton, NRAO/AUI/NSF)**

The explosion of the 'ordinary' supernova of this type, known as a **core-collapse supernova**, causes a spherical blast wave of material to be sent into interstellar space. These jets can produce narrow beams of gamma rays, resulting in a gamma-ray burst, if in addition to this, a rotating disc of material briefly forms around the neutron star or black hole left behind the explosion and accelerates narrow jets of material at nearly the speed of light outward in opposite directions. The intense visible-light burst occurs shortly after the explosion that initially gave these objects their distinctive appearance when the dense material close to the star is impacted by the blast wave. Astronomers refer to these explosions as **"fast blue optical transients"** because of that dazzling flash as well. This is one among the traits that set them apart from common supernovae. Astronomers examine a variety of transient occurrences through radio observations, such as the explosive brightness of gamma-ray bursts (GRBs) and the enticing mysteries of fast radio bursts (FRBs). These occurrences provide a unique window into the cataclysmic processes taking place throughout the cosmos, revealing celestial dynamics and extreme physics that were previously unobservable.

### 9.2 Cosmic Explosions and Pulsar Timing

As cosmic explosions, fast radio bursts (FRBs) and gamma-ray bursts (GRBs) take centre stage among the mysteries of the cosmos that transient radio astronomy has unlocked. Astronomers are fascinated by the transient brightness of FRBs, which are observed as powerful, millisecond bursts of radio waves from unknown sources. Similar to this, GRBs, the cosmologically most intense explosions, emit enormous amounts of energy throughout the electromagnetic spectrum. Scientists are able to see into the heart of cataclysmic events that change galaxies and the cosmos itself through radio studies, helping them to understand the origins of these cosmic pyrotechnics. Furthermore, the precise timing of pulsars, which are neutron stars that spin quickly, is crucial for spotting minute alterations like the presence of gravitational waves. The most profound mysteries of the universe can be seen via a unique window provided by pulsar timing **[23]**, which reveals the secrets of fundamental physics.

### 9.3 Cosmological Insights and the Hubble Constant

Transient radio astronomy opens a door to cosmic ideas, revealing the universe's dynamic nature. Transient event research provides essential insights into the dispersion of matter, cosmic evolution, and the vastness of space. Astronomers map the cosmic web of large-scale structures and the expansion of the cosmos by studying transient occurrences at various distances and redshifts. Our grasp of fundamental ideas like dark matter, dark energy, and the cosmic microwave background radiation is aided by these cosmological revelations. Transient radio astronomy is also essential for determining the Hubble constant, a crucial factor in the universe's expansion rate. The accurate calculation of the Hubble constant **[24]** affects our understanding of the age and eventual destiny of the universe and places fundamental restrictions on cosmological models.

The cosmos' fleeting glories are illuminated by transient radio astronomy, which also provides profound insights into the cosmos' enormous and dynamic geography. We examine the universe's dynamic fabric through the use of radio telescopes, which helps us better understand cosmic events and leads us to a greater appreciation of the cosmos.

### 10. FUTURE DIRECTIONS

The potential for radio astronomy in the future is really intriguing since the cosmic curtain is still being peeled back. Projects and missions in radio astronomy, both current and planned, have the potential to expand our understanding of the universe. Astronomers are able to see deeper into the cosmos with previously unheard-of precision and sensitivity thanks to developments in technology and instrumentation. Future developments in the field have tantalising potential, with the possibility to solve even more complex mysteries that are concealed in the depths of space. The most puzzling facets of the cosmos may be uncovered by current and future radio astronomy initiatives. One of the most ambitious projects in the history of radio astronomy, the Square Kilometre Array (SKA), is poised to revolutionise our understanding of the cosmos. With a collecting area of one square kilometre spanning two continents, Australia and South Africa, the SKA will provide unprecedented sensitivity and resolution. This massive collection of radio telescopes will look into the deepest recesses of the cosmos and shed light on perplexing fast radio bursts (FRBs), the nature of dark matter, and other cosmic phenomena. Here is a list of some of the notable radio observatories in the world:

1. **Arecibo Observatory - Puerto Rico, USA** (Note: Arecibo Observatory was decommissioned in 2020)
2. **Green Bank Observatory - West Virginia, USA**
3. **Very Large Array (VLA) - New Mexico, USA**
4. **Parkes Observatory - New South Wales, Australia**
5. **Jodrell Bank Observatory - Cheshire, United Kingdom**
6. **Atacama Large Millimeter Array (ALMA) - Atacama Desert, Chile**
7. **Effelsberg 100m Radio Telescope - Effelsberg, Germany**
8. **Westerbork Synthesis Radio Telescope - Dwingeloo, Netherlands**
9. **Giant Metrewave Radio Telescope (GMRT) - Pune, India**
10. **Medicina Radio Observatory - Medicina, Italy**

11. Owens Valley Radio Observatory (OVRO) - California, USA
12. Dominion Radio Astrophysical Observatory (DRAO) - British Columbia, Canada
13. Nançay Radio Telescope - Nançay, France
14. Molonglo Observatory Synthesis Telescope (MOST) - New South Wales, Australia
15. Yebes Observatory - Yebes, Spain

This list is not comprehensive; there are numerous additional radio observatories located all over the world that each conduct important research in the field of radio astronomy. Additionally, some observatories may be decommissioned or repurposed over time, while others might be built specifically to study the cosmos using radio waves.

Radio astronomy is being transformed by developments in technology and instruments. Weak radio signals from far-off astronomical objects are becoming easier to detect and investigate thanks to new receiver technologies and data processing methods. We can expect to learn more about the radio sky **[25]** as more sensitive radio detectors, such cryogenically cooled receivers and phased array feeds, are developed. The rapid and effective processing of massive data sets made possible by developments in computing and data analysis is another factor that makes it possible to glean important scientific insights from the vast amounts of radio wave data amassed.

Future developments in radio astronomy have limitless possibilities. The field is poised to make important discoveries, from figuring out the enigmas of dark energy and dark matter to learning more about exoplanets and black holes. Astronomers all over the world are still motivated by the pursuit of knowledge regarding the characteristics of the early universe, the existence of extraterrestrial life, and the fundamental principles regulating the cosmos. The possibility of revealing previously undiscovered cosmic events and upending accepted paradigms is ever-present as we go deeper into radio frequencies and investigate novel observational techniques.

The quest for deeper cosmic insights and the evolution of technology are driving several current and planned radio astronomy projects, which, in turn, throw a bright light on the field's bright future. Radio astronomy is set to alter our understanding of the cosmos and push the boundaries of human knowledge to new heights, especially with the SKA on the horizon and advancements in instrumentation.

11. **CONCLUSION**

We set out on an educational journey through the fascinating field of radio astronomy throughout this study article, revealing significant insights into the universe' mysteries. After highlighting the significance and scope of our inquiry in the introduction, we moved on to a historical overview where we saw significant turning points and the development of radio telescopes. On top of this foundation, we studied the principles of radio astronomy and gained an understanding of electromagnetic radiation's properties and connections to celestial objects. We investigated the complexities of radio telescope technology, as well as contemporary developments that support our cosmic research.

We dove into the field of radio astronomy observations, where a variety of tools and data-processing strategies unveiled celestial marvels. The mysterious Cosmic Microwave Background (CMB) radiation stood out among these discoveries, providing priceless insights into the early history of the universe. We travelled through the study of the Milky Way galaxy and beyond, exploring extragalactic spheres. While extragalactic investigations revealed the secrets of far-off galaxies and active galactic nuclei, galactic radio astronomy unlocked the mysteries of pulsars, supernova remnants, and neutral hydrogen maps. The transient radio astronomy segment gave us a peek of cosmological concepts like the Hubble Constant and the dynamical character of the universe while introducing us to the world of cosmic explosions and pulsar timing. We observed present and forthcoming studies, technological developments, and the potential for future ground-breaking discoveries as we looked into the future paths of radio astronomy.

This research paper has been a voyage of amazement and discovery, to sum it up. Our knowledge of the cosmos and our place within it has been shaped by the startling revelations made possible by radio astronomy. Radio astronomy has established itself as a crucial tool in our quest to fully grasp the scope of creation, from the interior of our Milky Way to the furthest reaches of distant galaxies.

As we say goodbye to our voyage, let us acknowledge the significance of radio astronomy as a doorway to knowledge and enlightenment. As a result of its continual developments and prospective discoveries, future generations will be motivated to continue delving into the universe' mysteries. Radio astronomy will continue to be our dependable guide on this ageless voyage of cosmic inquiry as the cosmos awaits, its wonders still to be revealed.